# Upper critical fields and critical current density of $BaFe_2(As_{0.68}P_{0.32})_2$ single crystal

**S.V. Chong,* S. Hashimoto, K. Kadowaki**

*Institute of Materials Science and Graduate School of Pure & Applied Sciences, University of Tsukuba, 1-1-1, Tennodai, Tsukuba, Ibaraki 305-8573, Japan*

* Corresponding author
E-mail address: s.chong@irl.cri.nz (S.V. Chong)
Present address: Industrial Research Ltd, P.O. Box 31310, Lower Hutt 5040, New Zealand

**Abstract**. The transport properties, upper critical fields, superconducting anisotropy, and critical current density of an iso-valent phosphorus-doped $BaFe_2As_2$ single crystal close to optimum doping are investigated in this report. Temperature dependent resistivity and susceptibility show a superconducting transition temperature, $T_c$, just below 31 K both with sharp transitions. The upper critical field parallel to the ab-plane, $H_{c2}^{ab}$, is above 77 Tesla while that along the c-axis direction, $H_{c2}^{c}$, is just above 36 Tesla, yielding a low superconducting anisotropy ratio ~ 2. The estimated inter-plane coherence length based on the Ginzburg-Landau (G-L) theory indicates $BaFe_2(As_{0.68}P_{0.32})_2$ is still above the point for a dimensional crossover inferring the superconducting layers are not weakly-coupled in this system. The critical current density at 5 K obtained from magnetization loops measurement show a modest $J_c$ as high as $10^5$ A/cm$^2$.

## 1. Introduction

Iron-arsenide based superconductors possess many interesting properties and behaviors which have conjured up a huge amount of interest in the superconductivity community since the initial breakthrough reported by Kamihara *et al.* [1]. Among them are the antiferromagnetic ordering and structural phase transition due to the spin-density wave (SDW) instability which was first observed in the rare-earth iron oxypnictides, later on also realized in the 122 iron-arsenides ($AFe_2As_2$, where $A$ = Ba, Sr, Ca, and Eu) [2-5]. These magneto-structural phase transitions are noticeable from several types of measurement, including temperature dependent resistivity, specific heat capacity, neutron-scattering, Mössbauer spectroscopy, and NMR measurements to list a few [6-10]. The suppression of these transitions by doping (either by electron or hole) in most cases induces superconductivity in the 122-FeAs system. This has been well demonstrated through hole doping with alkali metals (Na, K, or Cs) [11-15] where $T_c$ as high as 38 K have been achieved, while electron doping by substituting a small fraction of the Fe with larger transition metal elements such as Co, Ni, and more recently with Rh, Ru, Rb, Ir, and Pd [16-21] yield superconductivity with lower $T_c$; there are also cases where superconductivity was not realized even when this SDW anomaly is suppressed by doping [22]. Applying pressure (hydrostatic) to the 122-FeAs parent compounds can also suppress this anomaly and induces superconductivity with $T_c$ reaching up to 31 K in $BaFe_2As_2$, 34 K in $SrFe_2As_2$, 13 K in $CaFe_2As_2$, and 30 K in $EuFe_2As_2$ [23–26]. Chemical pressure is another novel route to induce superconductivity in the 122-FeAs system when phosphorus is substituted for arsenic in the FeAs layers. This was demonstrated first in $EuFe_2(As_{1-x}P_x)_2$ [27], and more recently in $BaFe_2(As_{1-x}P_x)_2$ with $T_c$ showing to be comparable to that of hydrostatic pressure induced superconductivity in $BaFe_2As_2$ peaking at around 31 K [28,29].

Recently, there have been several comparative studies on the fundamental properties between $BaFe_2(As_{1-x}P_x)_2$ and $Ba_{1-x}K_xFe_2As_2$ single crystals [30,31]. These studies reveal that although the two intermetallic superconducting compounds have comparable $T_c$, superconducting phase diagram, and similar Fermi surface topology, their superconducting energy gaps are surprisingly different – $BaFe_2(As_{1-x}P_x)_2$ has an energy gap with line nodes compared with the nodeless gap found in the $Ba_{1-x}K_xFe_2As_2$ [30]. Moreover, $^{31}$P NMR shows the nuclear spin-lattice relaxation rate $(T_1)^{-1}$ and Knight shift $^{31}$P $(T_1T)^{-1}$ behaviors of $BaFe_2(As_{0.67}P_{0.33})_2$ in the normal and superconducting state are distinctly different in many ways from the other iron-arsenides. The low temperature NMR data, supported by thermal conductivity measurements, infers the existence of a residual density-of-state (DOS) at zero energy in this phosphorus-doped superconductor [31].

The several reports to date on $BaFe_2(As_{1-x}P_x)_2$ single crystals have focused on the superconducting mechanisms of this system [29-32]. In this study we probe into some of the practical superconducting properties of a near optimally doped $BaFe_2(As_{1-x}P_x)_2$ single crystal, which have not been fully characterised in previous works [29,30]. The upper critical fields and the superconducting anisotropy were determined from magnetic fields dependence of the temperature dependent in-plane resistivity at two different orientations along the fundamental crystallographic axes. High upper critical fields but a low superconducting anisotropy are found in this compound. Based on the magnetization hysteresis loops in the superconducting state a high critical current density is obtained.

## 2. Experimental Details

Single crystals of $BaFe_2(As_{1-x}P_x)_2$ close to optimum doping were grown following a two-step method. First, a polycrystalline precursor was prepared by reacting Ba metal pieces with Fe,

As, and P powders in a 1.05:2:1.3:0.7 molar ratio placed inside an alumina crucible and sealed in an evacuated quartz tube. The assembly was first heated slowly to 600 K, held for 12 h, then the temperature was raised slowly again to 1173 K and held for 40 h following the procedure in ref. [28]. The sintered precursor was thoroughly ground into fine powder and packed into a ~4 mm inner diameter alumina u-tube, which was then placed inside a screw-cap niobium container. The niobium container was placed inside an inductive furnace filled with partially reduced purified argon gas, heated to well above 1500 K until melting occurs and then cooled at a rate of 1 K/min. to 1153 K before the furnace was switched off to allow rapid cooling to room temperature. Platelet crystals as large as $1 \times 1 \times 0.05$ mm$^3$ could easily be extracted with sharp c-axis (00*l*) Bragg reflections in the x-ray diffraction (XRD) scans (Fig. 1). The composition of phosphorus was determined from energy-dispersive x-ray microanalysis spectrum (EDS) giving a normalized composition of $BaFe_2(As_{0.68}P_{0.32})_2$, which is close to the optimally doped composition [28,29]. The temperature dependent susceptibility and field dependent magnetization were measured on a Quantum Design SQUID magnetometer with maximum applied magnetic fields of ± 5 Tesla. Temperature dependent resistivity measurements under magnetic fields up to 5 Tesla were conducted on a Quantum Design physical property measurement system (PPMS); a separate standard four-probe dc method was also used to study the temperature dependent resistivity (*R*-*T*) behaviour of the single crystals at zero field.

### 3. Results and Discussion

Figure 2 presents the temperature dependent in-plane resistivity ($\rho_{ab}$) of $BaFe_2(As_{0.68}P_{0.32})_2$ single crystal measured at zero magnetic field. The SDW anomaly occurs just below 140 K as observed in the parent compound, $BaFe_2As_2$, is completely suppressed and a linear $\rho(T)$ behavior is observed below 150 K down to $T_c$, which resembles those of near optimally doped $BaFe_2(As_{1-x}P_x)_2$ single crystals and polycrystalline samples [28,29]. Kasahara *et al.* have also studied the Hall effect of near optimally P-doped Ba-122 single crystals in which they found a strong temperature dependent Hall coefficient ($R_H$) and the magnitude of $R_H$ at low temperature is several times larger than that predicted from a multiband theory. From this and the nearly perfect $T$-linear $\rho(T)$ in the normal state resistivity the authors point to non-Fermi liquid transport properties similar to those observed in strongly correlated electron systems [29]. At low temperature, our sample shows a sharp resistive superconducting transition at $T_c$ onset $\approx$ 30.9 K with a superconducting transition width $\Delta T_c \approx$ 0.9 K between the 90% and 10% points of resistive transition. The residual resistivity at 31 K is about 22 μΩ·cm with a residual resistivity ratio (*RRR*) defined by ρ(300 K)/ρ(31 K) of ~ 7.7. Both XRD and transport property results confirm this sample is close to optimum doping and is of good quality.

Temperature dependent susceptibility (*M-T*) for a $BaFe_2(As_{0.68}P_{0.32})_2$ single crystal measured at low fields with $H \parallel$ c-axis indicates a similarly sharp diamagnetic transition at 30.1 K which is close to the $T_c$ mid-point at 30.2 K from *R-T* measurement (inset of Fig. 2). The higher field *M-T* of another piece of $BaFe_2(As_{0.68}P_{0.32})_2$ single crystal is shown in Fig. 3 for a magnetic field applied along the c-axis ($H \parallel$ c) and parallel to the ab-plane ($H \parallel$ ab). The behaviour of the zero-field-cooled (*ZFC*) and field-cooled (*FC*) data are similar to the data obtained in low field, with similar $T_c$, and also for the two field orientations. An estimation of the superconducting volume fraction from the Meissner effect's *FC* data for a field applied (20 Oe) along the ab-plane is about 1.6 % and the corresponding diamagnetic shielding is 16 % confirming bulk nature of the superconductivity.

Estimations of the upper critical fields ($H_{c2}$) of $BaFe_2(As_{0.68}P_{0.32})_2$ single crystal were performed by measuring the temperature dependent in-plane resistive ($\rho_{ab}$) at magnetic fields up to 5 Tesla applied along the c-axis and ab-plane, respectively. As shown in Fig. 4 the superconducting transitions are slightly broadened and the resistive transition curves shift parallel down to lower temperatures with increasing applied fields, which indicates high upper critical fields. Moreover, the broadening and the resistive shift are slightly larger for a magnetic field applied along the c-axis than along the ab-plane, indicating an anisotropy ratio larger than unity. The inset in figure 4a shows the irreversibility ($H_{irr}$) and upper critical fields, defined by the 10 % and 90 % points of the resistive transition curves, for the two different orientations plotted as a function of temperature. The curves of $H_{c2}(T)$ show average slopes $-dH_{c2}/dT|_{Tc}$ = 3.66 Tesla/K for $H$ || ab and 1.72 Tesla/K for $H$ || c. According to the Werthamer-Helfand-Hohenberg (WHH) formula [33], $H_{c2}(0) = -0.69(dH_{c2}/dT|_{Tc})\cdot T_c$, and using $T_c$ = 30.6 K (at 90% of $\rho(T_c)$), the value of upper critical fields at zero temperature are $H_{c2}(0)$ = 77.4 Tesla along the ab-plane and 36.4 Tesla along the c-axis, giving an upper critical field anisotropy ratio $\gamma_H = H_{c2}^{ab}/H_{c2}^{c}$ = 2.13. A similar anisotropy ratio is also obtained from the irreversible field with $H_{irr}^{ab}/H_{irr}^{c}$ = 2.07. The value of upper critical field here is slightly higher than that reported by Hashimoto *et al.* ($H_{c2}$ ~ 52 Tesla) from specific heat data for a $BaFe_2(As_{0.67}P_{0.33})_2$ single crystal [30]. Table 1 lists the upper critical fields and anisotropy values of several 122 and 111-FeAs superconducting single crystals reported in literature. In general the hole-doped systems have higher upper critical fields > 100 Tesla compared to the electron-doped system, while those of strain or pressure induced superconductivity have similar upper critical fields as in the electron-doped system. Except for $KFe_2As_2$, the anisotropy ratios of the non-oxygen containing FeAs-based superconductors are lower than 3.6, which is in contrast to the 1111 iron oxypnictides with anisotropy ratios larger than 4 [43].

Based on the Ginzburg-Landau (G-L) theory, estimations of the coherence lengths just below $T_c$ at 29 K were also carried out to investigate the superconducting dimensionality for the in-plane [$\xi_{||ab} = (\phi_0/2\pi H_{c2}^{c})^{1/2}$] and inter-plane [$\xi_{||c} = \phi_0/2\pi(\xi_{||ab})H_{c2}^{ab}$] directions (where $\phi_0$ is the magnetic flux quantum = $2.07 \times 10^{-7}$ Oe·cm$^2$; estimated $H_{c2}^{ab}$(29 K) ≈ 79 Tesla; $H_{c2}^{c}$(29

K) ≈ 32 Tesla). The calculations yield $\xi_{||ab} \approx 32$ Å and $\xi_{||c} \approx 13$ Å, where the latter inter-plane $\xi$ is still longer than the c-axis spacing of $BaFe_2(As_{0.68}P_{0.32})_2$ of ca. 12.8 Å, indicating the superconducting layers are not decoupled. This finding is in-line with the conclusions derived from the studies on $(Ba_{1-x}K_x)Fe_2As_2$ and $Ba(Fe_{1-x}Co_x)_2As_2$ single crystals at high magnetic fields which also indicate a three-dimensional nature of the superconductivity in 122-FeAs system [36,38]. It should be noted that $\xi_{||c}$ from our study is about two times smaller than those of potassium and cobalt-doped Ba-122 samples mentioned above, which might in some ways add to the observed differences in the superconducting DOS state of $BaFe_2(As_{1-x}P_x)_2$ and the nesting conditions in the Fermi surfaces obtained from NMR study compared with other iron-arsenide superconductors [31].

The isothermal superconducting hysteresis loops (*M–H*) at various temperatures below $T_c$ up to ± 5 Tesla applied magnetic fields parallel to the c-axis of the single crystal are shown in Fig. 5a. At the high field ends irreversible magnetization is clearly seen below 15 K but disappears at 25 K. An estimation of the critical current density ($J_c$) was carried out by assuming a homogenous current flows within the sample using the Bean model [44], $J_c = 20(M^- - M^+)/a(1 - a/3b)$ where $M^-$ and $M^+$ (unit: emu/cm$^3$) are the magnetization when sweeping the fields down and up, respectively; *a* and *b* (unit: cm) are the dimensions of the sample with a < b. At low fields, $J_c$ exhibits a rapid decrease at first but becomes weakly field dependent at high fields (Fig. 5b). For temperatures below 15 K, $J_c$ is above $10^4$ A/cm$^2$ even up to fields of 5 Tesla in this single crystalline sample. The self-field $J_c$ at 5 K is as high as ~5 × $10^5$ A/cm$^2$ which is similar in range to that obtained from other Ba-122 superconducting single crystals [20,37,45]. Indeed all the Ba-122 superconducting single crystals, whether they are hole or electron doped, show high critical current density. For example, $Ba_{0.6}K_{0.4}Fe_2As_2$ has self-field Jc at 2 K of more than 4 × $10^6$ A/cm$^2$ [45] and self-field Jc ~ 4 × $10^5$ A/cm$^2$ at 4.2 K in $Ba(Fe_{0.9}Co_{0.1})_2As_2$ single crystal [37]. The high Jc in these two compounds has been attributed to the distortion of the dopant (K and Co) in their lattice sites. The Jc values in our crystal are slightly lower than those mentioned above in comparable field and temperature ranges, indicating less distortion in our clean sample, which is as expected when arsenic is substituted by phosphorus in $BaFe_2(As_{1-x}P_x)_2$ [28,29]. Moreover, our crystals are not post-annealed and some of the atoms are likely to be distorted from their ideal lattice positions inducing some pinning effect.

## 4. Summary

The transport properties, anisotropic of upper critical field, and critical current density have been studied for a phosphorus-doped $BaFe_2As_2$ single crystal close to optimum doping. Temperature dependent resistivity shows a near-linear temperature dependent behavior close to and below the suppressed spin-density wave anomaly (< 150 K) supporting the non-Fermi liquid transport properties observed in the previously reported $BaFe_2(As_{0.67}P_{0.33})_2$ single crystal [29]. The irreversibility and upper critical fields are quite similar with values > 32 Tesla and a weak anisotropy $\gamma \sim 2$. From the estimated inter-layer coherence length, it is concluded that superconductivity in P-doped $BaFe_2As_2$ is not weakly coupled two dimensionally but three dimensional in nature similar to the charge-doped counterparts. Field dependent magnetization measurements show hysteretic loops behaviour with observable irreversible magnetization at the high field ends up to 15 K. The critical current density calculated from the Bean critical state model yields a respectable $J_c$ in the range between $10^4$ to $10^5$ A/cm$^2$ for temperatures below 15 K up to magnetic fields of 5 Tesla.

## Acknowledgements

This work has been supported by JSPS-KAKENHI, Grant-in-Aid for Scientific Research (A) (18204031), the Ministry of Education, Culture, Sports, Science and Technology (MEXT), JAPAN, CREST JST, WPI at NIMS (MANA), JSPS Core-to-Core Program - Strategic Research Networks, “Nanoscience and Engineering in Superconductivity (NES)”, and the JSPS postdoctoral fellowship for foreign researchers.

**Table 1**. Superconducting parameters of 122 and 111-iron-arsenides single crystals.

| Sample | $T_c$ (K) | $H_{c2}^{ab}$ (T) | $H_{c2}^{c}$ (T) | $\gamma_H$ | Ref. |
|---|---|---|---|---|---|
| $BaFe_2As_2$ | 22.5 | ~88 | ~31 | ~2.8 | [34] |
| $Ba_{0.6}K_{0.4}Fe_2As_2$ | 36.5 | 235 | 135 | 2 – 3 | [35] |
| $Ba_{0.55}K_{0.45}Fe_2As_2$ | 30 | ≥75 | ≥75 | 1.2 – 3.5 | [36] |
| $Ba_{0.84}Rb_{0.10}Sn_{0.09}Fe_2As_2$ | 23 | 120 | 70 | 2.4 – 3 | [20] |
| $Ba(Fe_{0.9}Co_{0.1})_2As_2$ | 22 | ~70 | ~50 | 1.5 – 2 | [37] |
| $Ba(Fe_{0.92}Co_{0.08})_2As_2$ | 23 | 55 | 50 | 1.1 – 3.4 | [38] |
| $SrFe_2As_2$ | 21 | ~70 | ~24 | ~2.6 | [39] |
| $SrFe_2As_2$ (pressure) | 34.1 | 86 | – | – | [24] |
| $Sr_{0.6}K_{0.4}Fe_2As_2$ | 35.5 | 185.4 | 93.1 | 2.0 | [40] |
| $CaFe_2As_2$ (pressure) | 12 | 20 | 15 | ~1.3 | [25[b]] |
| $KFe_2As_2$ | 2.79 | 4.47 | 1.25 | 3.5 – 6.8 | [41] |
| $Na_{1-\delta}FeAs$ | 15 | 60 | 33 | ~1.8 | [42] |

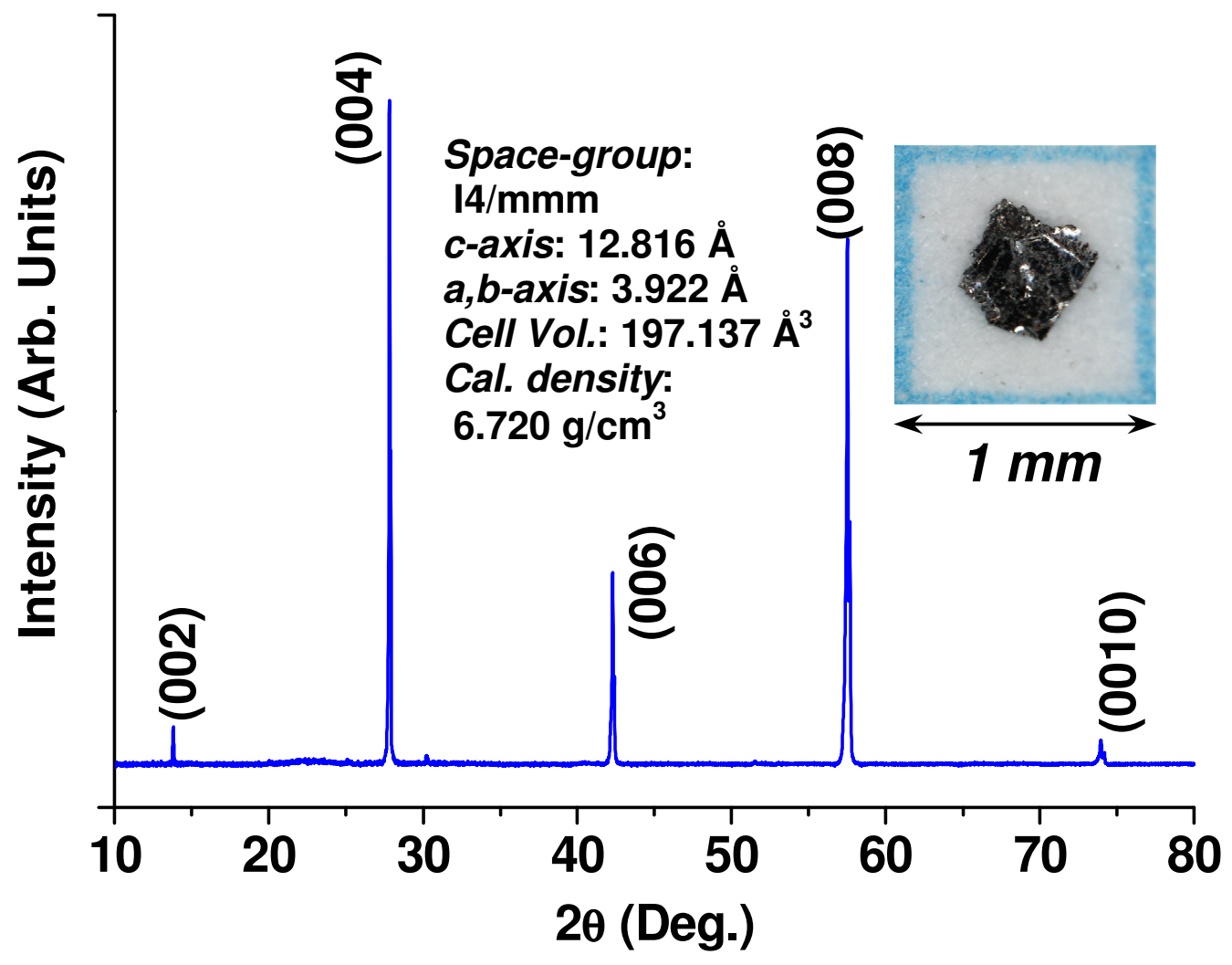

**Fig. 1**. X-ray diffraction pattern of $BaFe_2(As_{0.68}P_{0.32})_2$ single crystal. The lattice parameters were obtained from indexing of a powder XRD pattern of crushed single crystals.

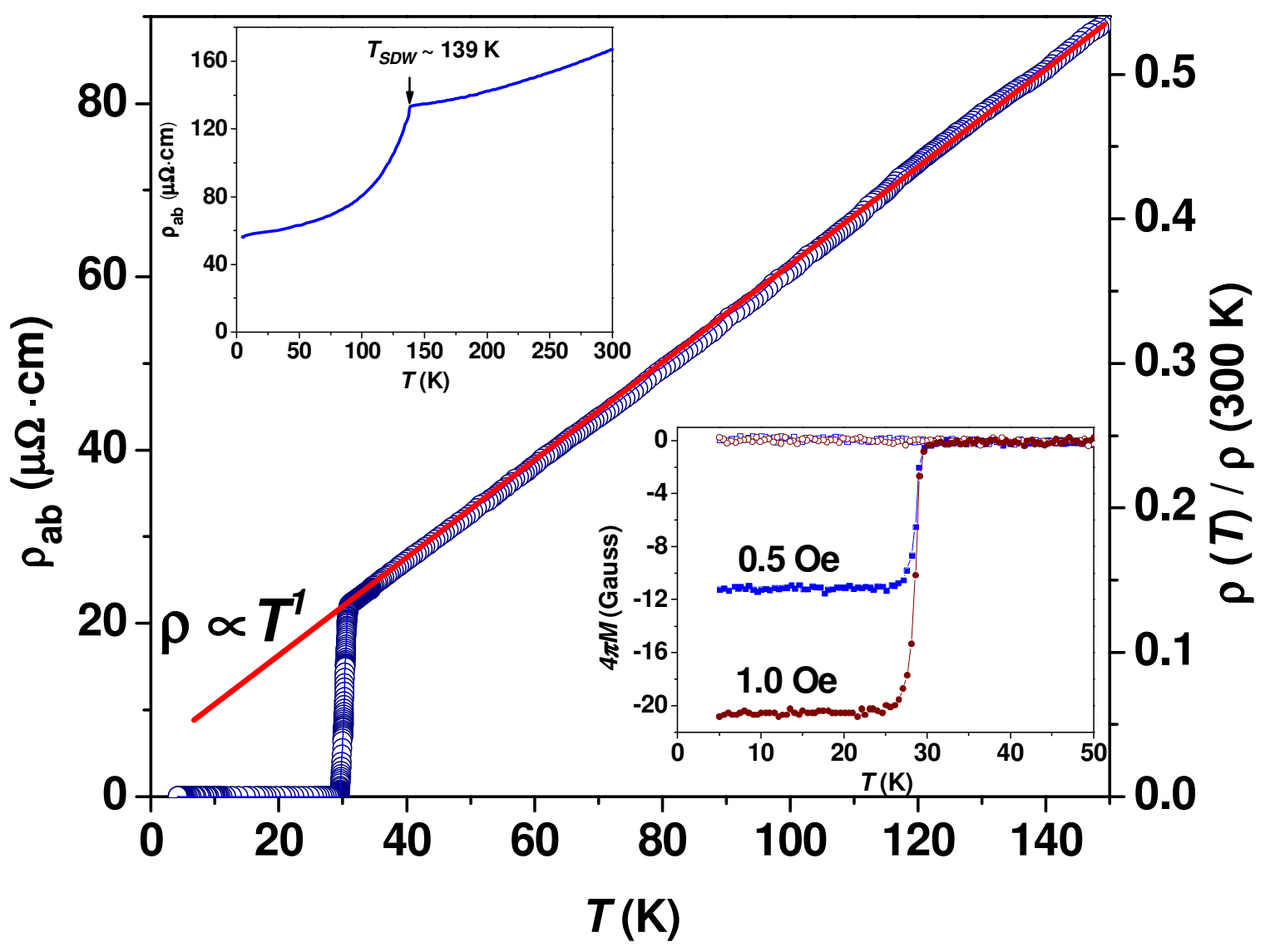

**Fig. 2**. Temperature dependent resistivity and susceptibility (inset: right bottom) of $BaFe_2(As_{0.68}P_{0.32})_2$ single crystal. The *R*-*T* of a non-doped $BaFe_2As_2$ single crystal grown from self-flux method is also shown (inset: top left) for comparison.

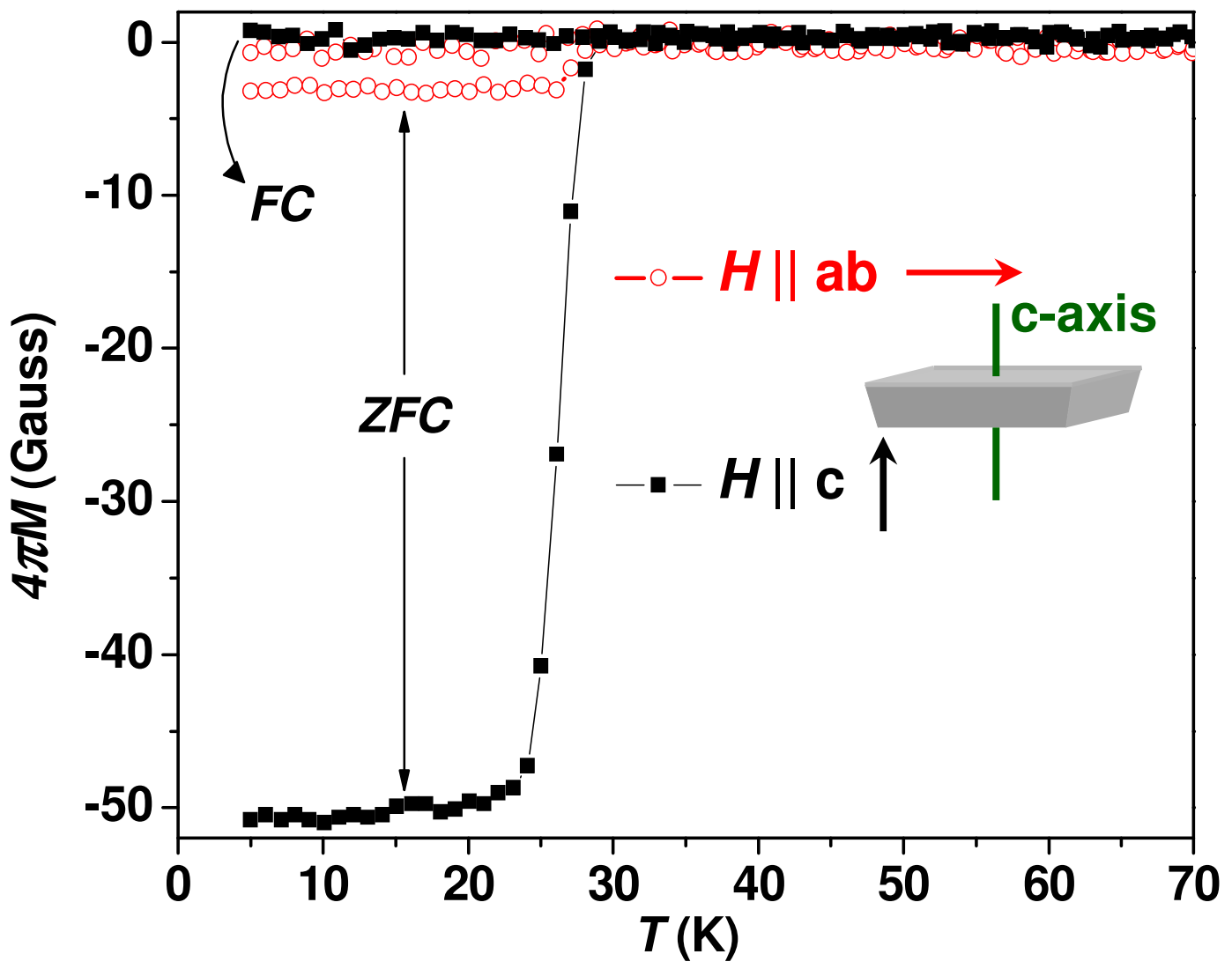

**Fig. 3**. Temperature dependence of the susceptibility measured at 20 Oe with the magnetic field applied parallel to the ab-plane (*H* || ab) and to the c-axis (*H* || c) as depicted schematically in the inset.

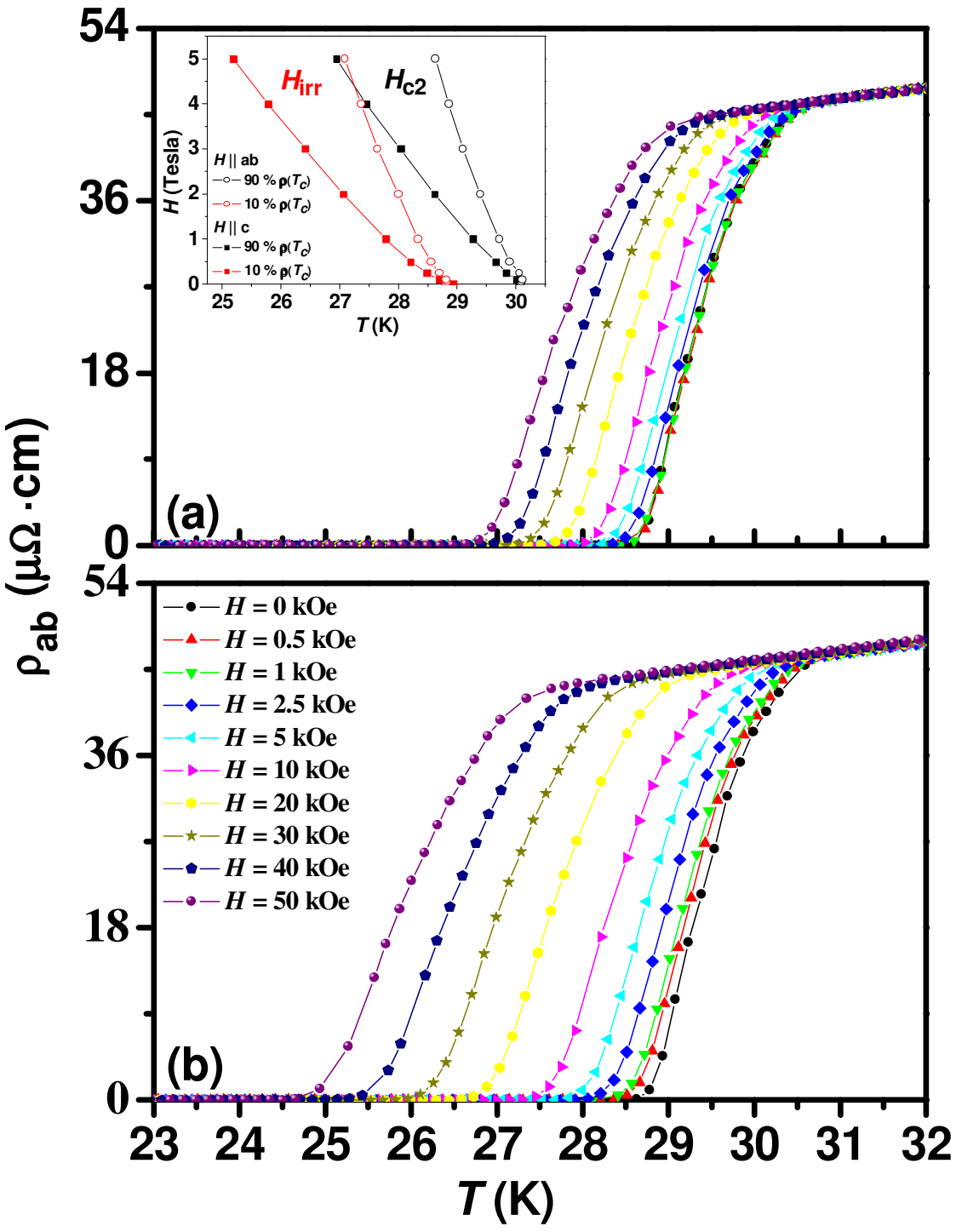

**Fig. 4**. In-plane resistivity versus temperature at various magnetic fields of $BaFe_2(As_{0.68}P_{0.32})_2$ single crystal measured with 0.2 mA applied current. The applied magnetic fields are (a) parallel to the ab-plane and (b) parallel to the c-axis. The inset in (a) shows the derived temperature dependent irreversibility ($H_{irr}$) and upper critical fields ($H_{c2}$) in both orientations.

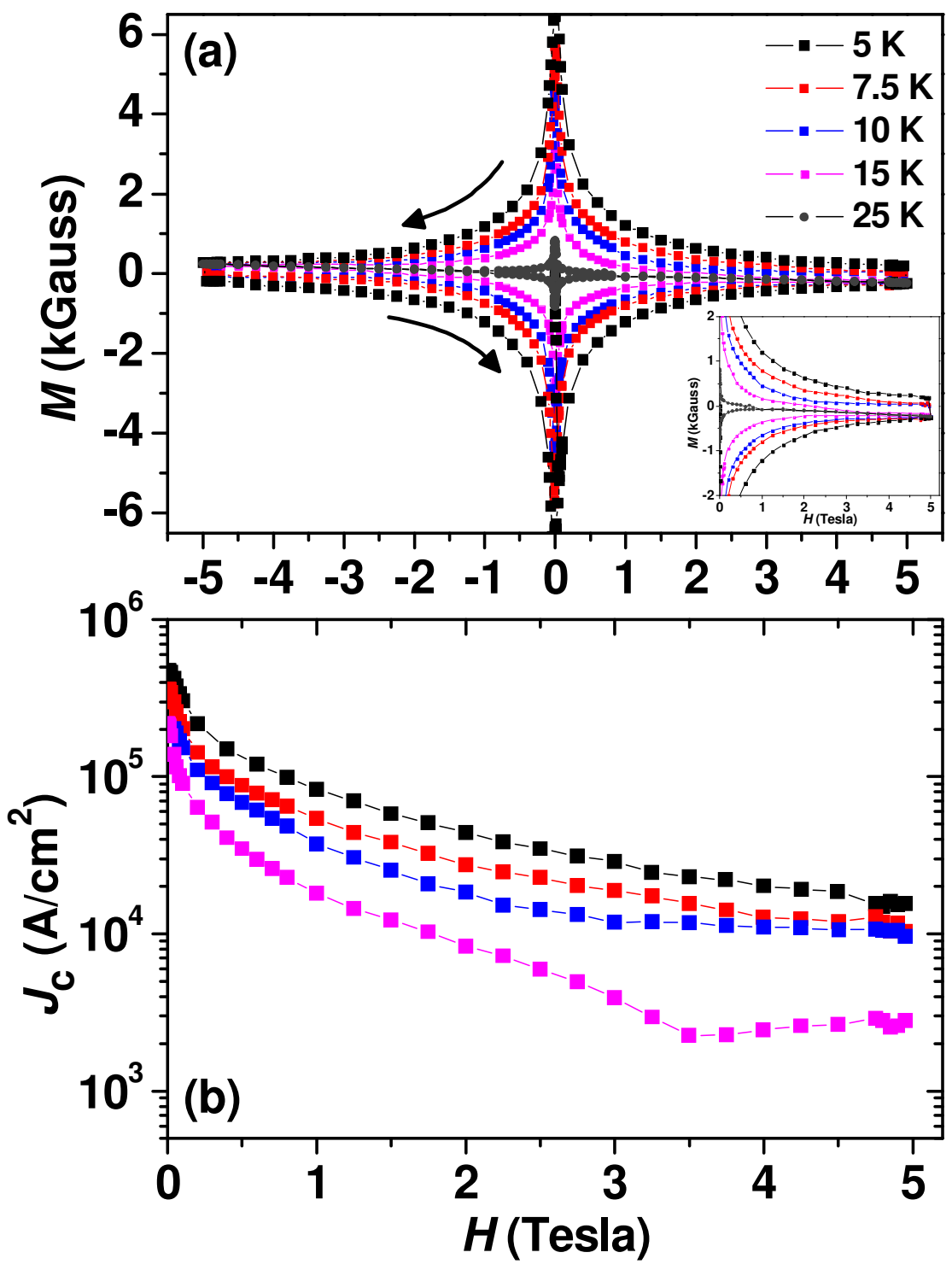

**Fig. 5**. Field dependence of magnetization for a $BaFe_2(As_{0.68}P_{0.32})_2$ single crystal with sample edge dimensions *a* ~ 0.63 mm and *b* ~ 0.81 mm (a) – the inset shows an enlargement of the positive field region of the *M*-*H* plot. The corresponding field dependence of critical current density below 25 K are shown in (b).